\documentclass[letterpaper]{article} 
\usepackage{aaai2026} 
\nocopyright
\usepackage{times}  
\usepackage{helvet}  
\usepackage{courier}  
\usepackage[hyphens]{url}  
\usepackage{graphicx} 
\usepackage{natbib}  
\usepackage{caption} 
\usepackage{algorithm}
\usepackage{algorithmic}
\usepackage{placeins} 
\usepackage{newfloat}
\usepackage{listings}
\DeclareCaptionStyle{ruled}{labelfont=normalfont,labelsep=colon,strut=off}
\floatstyle{ruled}
\newfloat{listing}{tb}{lst}{}
\floatname{listing}{Listing}

\usepackage{booktabs}

\usepackage{xcolor} 

\title{Food as Soft Power: Taiwanese Gastrodiplomacy on Social Media \\
and Algorithmic Suppression}

\author{
  Andrew Yen Chang\textsuperscript{1,2}\thanks{These authors contributed equally to this work.} \quad
  Ho-Chun Herbert Chang\textsuperscript{3}\footnotemark[1] \\[0.5em]
  \textsuperscript{1}University of California, Berkeley \\
  \textsuperscript{2}Institut d'\'{e}tudes politiques de Paris (Sciences Po Paris) \\
  \textsuperscript{3} Program in Quantitative Social Science, Dartmouth College \\[0.5em]
  {\normalsize\textcolor{gray}{\texttt{andrewchang382@berkeley.edu}, \texttt{herbert.chang@dartmouth.edu}}}
}

\begin{document}
\maketitle

\begin{abstract}
Social media platforms have become pivotal for projecting national identity and soft power in an increasingly digital world. This study examines the digital manifestation of Taiwanese gastrodiplomacy by focusing on \textit{bubble tea}---a culturally iconic beverage---leveraging a dataset comprising 107,169 posts from popular lifestyle social media platform Instagram, including 315,279,227 engagements, 4,756,320 comments, and 8,097,260,651 views over five full years (2020--2024), we investigate how social media facilitates discussion about Taiwanese cuisine and contributes to Taiwan's digital soft power. Our analysis reveals that bubble tea consistently emerges as the dominant representation of Taiwanese cuisine across Meta's Instagram channels. However, this dominance also indicates vulnerability in gastrodiplomatic strategy compared to other countries. Additionally, we find evidence that Instagram suppresses bubble tea posts mentioning Taiwan by $1{,}200\%$---roughly a twelve-fold decrease in exposure---relative to posts without such mentions. Crucially, we observe a significant drop in the number of posts, views, and engagement following Lai’s inauguration in May 2024. This study ultimately contributes to understanding how digital platforms can enable or disable gastrodiplomacy, soft power, and cultural diplomacy while highlighting the need for greater algorithmic transparency. By noting Taiwan's bubble tea's digital engagement and footprint, critical insights are brought for nations seeking to leverage soft power through gastronomic means in a politicized digital era and researchers trying to better understand algorithmic suppression.
\end{abstract}

\section{Introduction}
On January 19, 2023, Google commemorated bubble tea with a caption stating,
``Originating in Taiwan, this cold drink usually consists of tea, milk, and tapioca pearls.
Bubble tea gained such popularity globally that it [is] officially announced as a new emoji on this day''
\citep{Google2023}.
Since 2010, bubble tea has achieved global prominence, with Taiwanese immigrants establishing shops in Hong Kong,
the United States, and the United Kingdom, and gradually in Dubai, Manila, and Mumbai---even in regions with
smaller Taiwanese diaspora communities \citep{TaiwanToday2011}.
Given Taiwan’s contentious history with China, the increasing visibility of bubble tea has taken on symbolic
significance, serving as a marker of Taiwanese sovereignty both physically and, increasingly, in digital spaces.

Gastrodiplomacy refers to the strategic deployment of a nation’s culinary heritage to foster cross-cultural
understanding and enhance its international image.
As a form of soft power---one that seeks to win hearts and minds through cultural appeal rather than coercion---
gastrodiplomacy employs food as a diplomatic tool to build bridges and generate goodwill across borders
\citep{Rockower2011}.
Within this framework, bubble tea has emerged as an iconic symbol in online discussions of Taiwan,
encapsulating the country’s innovative culinary culture and distinct national identity,
while reinforcing its broader soft power strategies on the global stage.

Globally, news consumption increasingly occurs online, and according to Pew Research, more than half of young
American adults---54\%---now obtain their news from social media \citep{Pew2024}.
Instagram, YouTube, and Facebook are the most influential platforms, ranked first, second, and third,
respectively, for social media news consumption.
Visual-first platforms appear especially well-suited for promoting lifestyle and culinary content,
facilitating affective connections between users and cultures.
This raises an important question: what role can social media play in advancing soft power?
Such potential is not without precedent---platforms have been weaponized to spread misinformation that
influences elections and international relations \citep{Chang2024,ChangWangFang2024}.
In response, companies such as Meta (owner of Facebook and Instagram) have sought to reduce users’ exposure
to political content in an effort to limit misinformation and its effects \citep{Treisman2024}.
However, what is deemed ``political'' can have direct consequences for Taiwan---an inherently political subject---
thereby complicating the relationship between platform policies and cultural diplomacy.

In this paper, we examine the intersection of Taiwan’s gastrodiplomacy and its most iconic culinary export,
bubble tea, to assess how users and algorithms on Instagram associate Taiwan with this symbol,
and whether such associations advance or hinder Taiwanese soft power.
Drawing on a dataset comprising 315{,}279{,}227 likes, 4{,}756{,}320 comments, and 8{,}097{,}260{,}651 views as
``algorithm and engagement metrics'' from Instagram, we investigate the performance of posts related to bubble tea
and Taiwan, and explore their broader implications for Taiwan’s digital influence.
Our analysis yields three principal findings.
First, bubble tea dominates mentions in Taiwanese food--related posts, far surpassing all other culinary references,
indicating a strong association between bubble tea and Taiwanese cuisine.
Second, paradoxically, posts about bubble tea that reference Taiwan receive lower levels of views, likes,
and comments---our defined ``algorithm and engagement metrics''---suggesting that this association may diminish
digital visibility.
Third, we find that Meta’s algorithms explicitly suppress the exposure (views) of bubble tea posts that reference Taiwan,
thereby constraining the effectiveness of Taiwanese soft power on these widely used platforms.
These findings carry important implications for understanding algorithmic transparency,
the suppression of content classified as ``political,'' and strategies for sustaining soft power
in the face of adverse algorithmic dynamics.

\section{Background and Contributions}

\subsection{Taiwan’s History of Gastrodiplomacy}
Gastrodiplomacy has long been a central component of Taiwan’s public diplomacy.
Multiple branches of government have invested in promoting Taiwan’s culinary culture abroad,
ranging from the Executive Yuan’s NT\$1.1 billion (approximately US\$34.08 million) investment between 2010 and 2013
to support the international expansion of the food service sector \citep{TaiwanToday2010}
to President Lai Ching-te’s 2024 initiative to elevate Taiwan’s global profile by showcasing local culinary and tourism assets
through cross-regional collaborations and targeted marketing campaigns \citep{RTI2024}.

Taiwanese cuisine is inherently politicized and often becomes a site of contested sovereignty.
For example, 85\textdegree{}C Bakery Cafe---a globally recognized Taiwanese chain---faced severe backlash in mainland China
after President Tsai Ing-wen visited one of its Los Angeles branches in August 2018 \citep{Kuo2018}.
In response, Chinese companies and netizens pressured food delivery apps such as Meituan-Dianping and Ele.me to delist the chain,
investors sold off shares in its parent company Gourmet Master (causing a 9.8\% drop in stock value),
and boycotts as well as cyberattacks targeted its Chinese website \citep{Kuo2018}.
More recently, in 2021, the Chinese government banned the importation of Taiwanese pineapples, wax apples, and pomelos---measures widely interpreted as retaliatory responses
to Taiwan’s diplomatic actions, such as hosting U.S.\ House Speaker Nancy Pelosi,
resulting in more than US\$20 million in trade losses \citep{Lu2022}.
These incidents illustrate how Taiwanese food producers can become entangled in the diplomatic struggle over Taiwan’s status,
irrespective of their own political intentions.
Consequently, Taiwanese cuisine carries the potential to serve as a symbol of a distinct ``Taiwanese identity''
that can simultaneously strengthen cultural pride and provoke political backlash.

By contrast, countries such as France and Japan have long leveraged culinary diversity as a soft power asset.
Japan promotes a wide array of dishes---including ramen, sushi, and yakitori---through initiatives such as
the global ``Cool Japan/Global Washoku'' campaign launched in 2006 \citep{Cabral2024}.
French cuisine is similarly central to its national image,
from the ubiquitous presence of French restaurants worldwide to the Elysée Palace’s use of haute cuisine in state dinners \citep{Barroux2022}.
French gastronomy has even been inscribed on UNESCO’s Intangible Cultural Heritage Lists since 2010 \citep{MEAE2010}.
More recently, South Korea has pursued a synergistic strategy,
combining its global entertainment exports (the ``Hallyu'' wave) with the promotion of traditional dishes
such as kimchi, ramyun, tteokbokki, and bibimbap \citep{Nihayati2022}.
Government initiatives have even enlisted K-pop groups to promote Korean cuisine internationally \citep{Mishan2022}.
Such examples highlight how culinary diversity can enhance gastrodiplomacy
and raise questions about the relative importance of variety and cultural breadth in sustaining soft power.

\subsection{Social Media as Forums of Diplomacy}
The rise of social media has extended the reach of Taiwanese gastrodiplomacy into digital arenas.
While Taiwanese cuisine shares many techniques, ingredients, and flavor profiles with Chinese and Japanese cooking \citep{Sui2015},
the Taiwanese government has consistently framed its culinary heritage as a unique cultural asset warranting promotion.
Social media now serves as a transnational public sphere capable of diffusing this distinct food culture.
Indeed, a large-scale study of over 1.8 billion English-language Facebook posts and 51 million Chinese-language Weibo posts
found that online conversations reflect the core--periphery structure predicted by World Systems Theory:
international relations are shaped not only by news coverage but also by patterns of public attention,
which in turn influence perceptions of foreign affairs \citep{Barnett2017}.
In this sense, social media platforms serve as vital channels for promoting Taiwanese food culture and,
by extension, advancing Taiwan’s soft power. Crucially, due to its shared history with China and alliance with the United States, Taiwan serves as an information buffer between autocratic and democratic states~\cite{weener2025beyond}. 

One notable instance of organic gastrodiplomacy is the emergence of the Milk Tea Alliance in 2020,
originally involving Taiwan, Hong Kong, and Thailand---each with its own distinct milk tea tradition.
The alliance originated from online disputes with Chinese users and bots,
eventually coalescing into the hashtag \#MilkTeaAlliance as a means of uniting voices opposing China’s political actions:
contesting its claims over Taiwan, resisting Hong Kong’s crackdown under the National Security Law,
and addressing political grievances in Thailand \citep{Schaffar2021}.
For Taiwan, the movement coincided with the aftermath of the 2020 presidential election,
when sovereignty and self-determination were central public concerns.
Similar cross-regional solidarity had surfaced earlier in Hong Kong’s 2014 Sunflower Movement,
encapsulated in the slogan ``Today’s Hong Kong, Tomorrow Taiwan,'' which gained renewed prominence after the 2020 security law \citep{Chang2021}.
Even after Hong Kong’s democratic freedoms were curtailed, solidarity networks persisted,
including ``Milk Tea Alliance -- Japan'' and the Taiwan Digital Diplomacy Association’s ``Milk Tea Alliance Project,''
designed to promote democracy and awareness of Taiwan.
These developments demonstrate how bubble tea, as a symbol,
became intertwined with democratic values through international digital activism.

\subsection{Social Media Mechanisms}
Given that social media constitutes a primary channel through which global audiences encounter lifestyle information,
its underlying mechanisms shape the form of cultural diplomacy and soft information warfare.
Within social media research, the ``algorithmic funnel'' framework offers a useful lens for conceptualizing content dissemination,comprising three stages: exposure, engagement, and algorithmic implications.

\textbf{Exposure} refers to the visibility of content in users’ feeds,
shaped by algorithms that evaluate factors such as engagement history, user preferences, content relevance, and recency \citep{Gunina2023}.
\textbf{Engagement} encompasses user interactions---most prominently, likes and comments---that signal content value.
Likes are relatively low-effort actions that indicate approval and may enhance visibility indirectly,
whereas comments represent a higher level of engagement that can more strongly influence algorithmic ranking \citep{Rockower2011}.
\textbf{Algorithmic amplification} occurs when platforms’ machine learning systems prioritize certain posts based on these signals,
often increasing the reach of content with higher engagement.
However, such processes can inadvertently favor sensational or misleading information,
contributing to the formation of echo chambers that limit diverse perspectives \citep{Flaxman2016}.

Historical cases illustrate both the democratizing potential and the risks of algorithm-driven communication.
During the Arab Spring, social media facilitated mobilization by enabling citizens to bypass state-controlled media,
coordinate mass gatherings, and broadcast events to global audiences \citep{Wolfsfeld2013,Smidi2017}.
In Egypt, for example, once users circumvented internet restrictions,
there was a surge in tweets tagged \#egypt, connecting domestic and international conversations on political developments \citep{Bruns2013}.
Similarly, the George Floyd protests in 2020 saw the hashtag \#JusticeforGeorgeFloyd go viral,
with widespread participation on Twitter and Facebook from diverse global audiences,
further amplifying the movement’s visibility \citep{Chang2022}.
These cases highlight social media’s potential to advance democratic movements through rapid, decentralized information sharing.

However, the same dynamics can exacerbate polarization and information manipulation.
The 2016 U.S.\ presidential election and the Brexit referendum demonstrated how foreign actors---most notably from Russia and China---
used bots, disinformation campaigns, and targeted narratives to deepen social divisions \citep{Francia2018,DelVicario2017,Ferrara2020}.
Domestically, misinformation regarding COVID-19 mask usage and the integrity of mail-in ballots,
largely circulated within conservative networks, influenced political discourse during the 2020 Democratic primaries \citep{Chen2021}.

In response to such concerns, Meta collaborated with academic researchers to examine algorithmic influence during the 2020 U.S.\ presidential election.
Findings suggest that altering exposure to different types of information had minimal impact on users’ political opinions or polarization levels,
underscoring the challenges of influencing attitudes within entrenched echo chambers \citep{Nyhan2023}.
Meta’s Third-Party Fact-Checking Program further revealed that misinformation was disproportionately concentrated in conservative groups
and was prevalent on Facebook News \citep{Gonzalez2023}.
Additional studies show that reducing social media usage ahead of elections did not significantly alter polarization or political views,
though it reduced the likelihood of questioning misleading content \citep{Allcott2024}.
Even presenting content in reverse-chronological order---thereby increasing exposure to political and low-credibility content---
had little measurable effect on political knowledge or attitudes \citep{Guess2023}.

While these findings illuminate the interplay between algorithms and political discourse in domestic contexts,
considerably less is known about their implications in international settings---particularly in the realm of shared democratic values among allied nations,
or in the domain of lifestyle-oriented platforms such as Instagram.

\subsection{Research Questions}
This study addresses two interrelated questions at the intersection of Taiwan’s diplomatic strategy
and the governance of global communication platforms.
First, we assess the viability of food as an instrument of soft power, including its potential strengths and vulnerabilities.
Second, we examine whether the corporations operating major social media platforms actively support democratic values,
given their expanding influence over political and cultural discourse.

Algorithmic amplification raises pressing concerns regarding transparency, accountability,
and the potential for misinformation---not only during high-stakes events such as elections and public health crises
but also in shaping the diffusion of Taiwanese soft power.
Depending on how platform algorithms weight exposure and engagement,
content associated with Taiwan may be promoted, suppressed, or rendered invisible to global audiences.
Such processes have implications for individual users’ autonomy in discovering content,
for public perceptions of Taiwan’s identity, and for the country’s broader international standing.

Based on these considerations, we pose the following research questions:
\begin{enumerate}
    \item Which types of Taiwanese food are most salient in shaping Taiwan’s international image,
          and how is attention distributed among them?
    \item Which critical events have defined the temporal evolution of Taiwanese gastronomy,
          and what has been their level of international exposure?
    \item To what extent does Instagram influence the visibility and engagement of posts related to Taiwan and gastrodiplomacy,
          specifically in terms of exposure (views) and engagement (likes, comments)?
\end{enumerate}

\section{Methodology}

\subsection{Data Collection and Curation}
The collection process for Meta’s data on posts related to Taiwanese cuisine was performed using the Meta Content Library,
and a full list of the keywords is provided in Appendix~A.
Searching for prominent words related to Taiwan’s cuisine, such as \#TaiwaneseFood, \#TaiwaneseCuisine, \#TaiwaneseEats,
and \#TaiwaneseNightMarket, yielded the posts used for this study.
The keywords were selected based on popular hashtags related to food and widely appreciated Taiwanese dishes.

The dataset used for this study included 107{,}169 posts and associated metadata such as the number of views, likes, comments,
URL links, and user IDs---specifically from Meta’s Instagram---that collectively garnered 315{,}279{,}227 likes,
4{,}756{,}320 comments, and 8{,}097{,}260{,}651 views.
The dataset ranges from December~31, 2019, to November~28, 2024.
To focus on international audiences, we limited the scope of this study to English-language posts,
as the goal was to analyze how Taiwanese soft power affects those outside the Chinese-speaking sphere,
and to non-business accounts, in order to observe how ordinary users---those most likely to consume Taiwan-related soft power messages---
naturally use and engage with the platforms.

\begin{figure}[t]
  \centering
  \includegraphics[width=\linewidth]{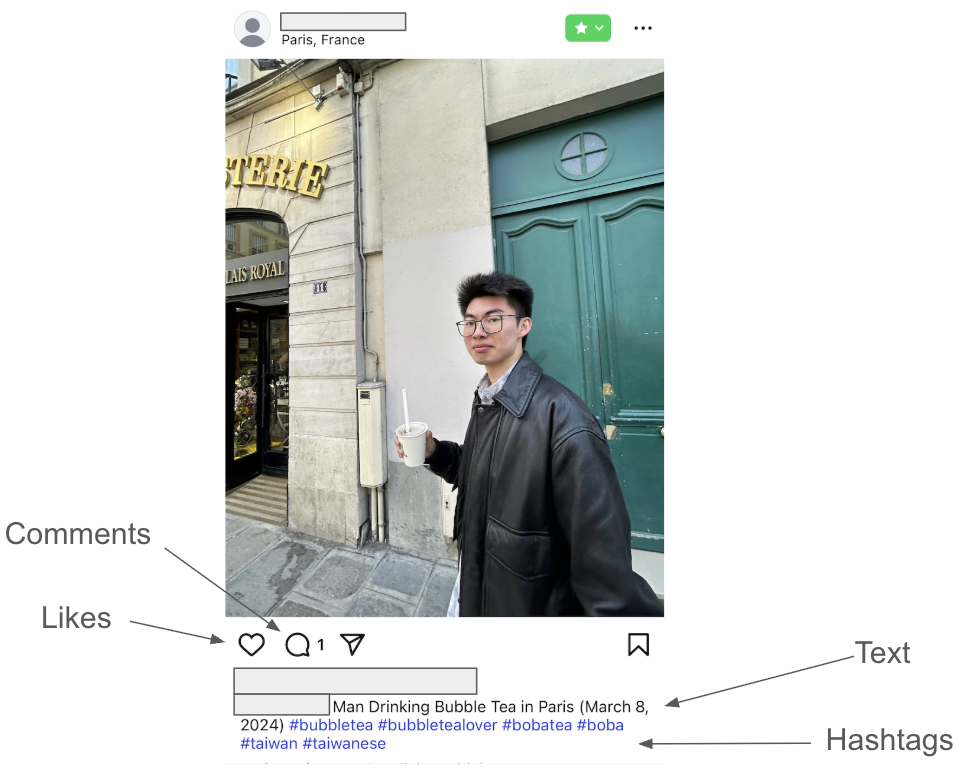}
  \caption{Parts of an Instagram post: example showing where likes, comments, text, and hashtags appear.}
  \label{fig:parts}
\end{figure}

\begin{figure}[t]
\centering
\includegraphics[width=\columnwidth]{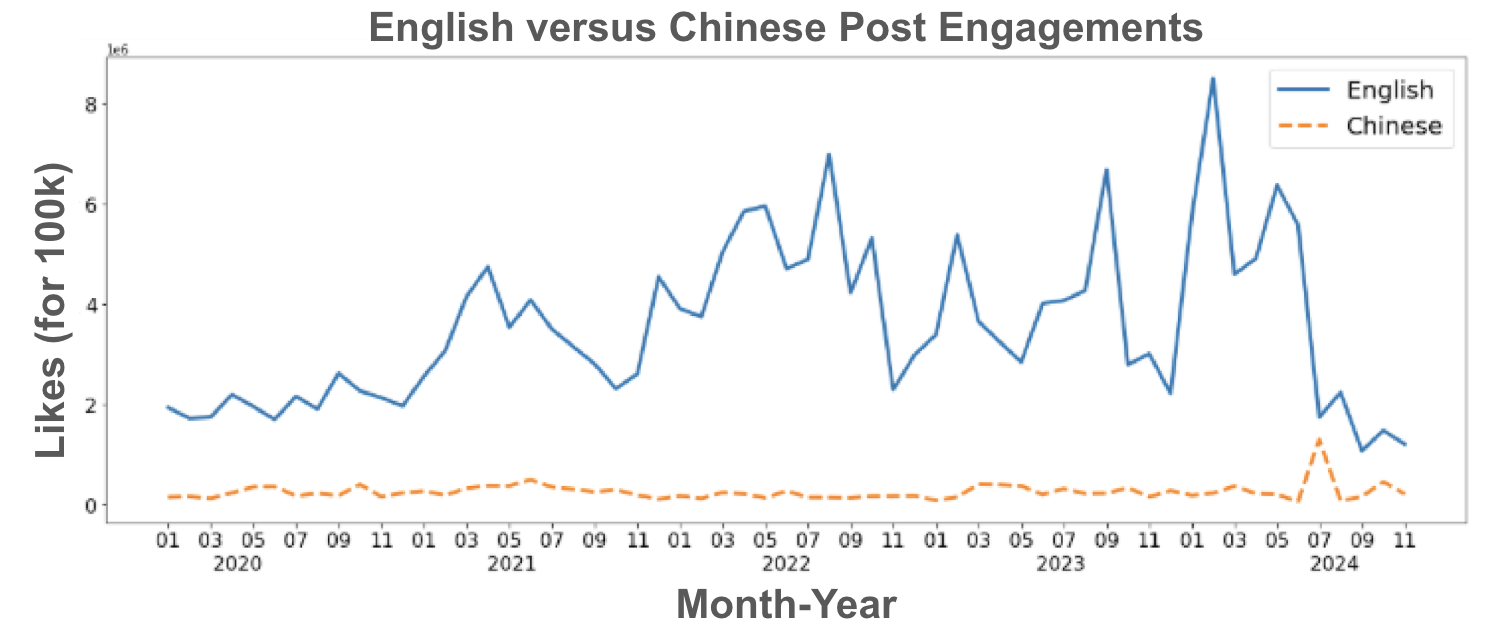}
\caption{Basic user engagement via likes of bubble tea posts (English versus Chinese) on Instagram 2020-2024..}
\label{fig:chinese_english}
\end{figure}

As part of a preliminary analysis, Figure 2 compares basic user engagement via likes
for bubble tea posts in English (solid blue line) versus Chinese (dashed orange line).
Since most foreign engagement outside the Sinosphere occurs in languages such as English,
focusing on English-language content provides a practical way to gauge how bubble tea---and, by extension, Taiwan---are perceived abroad.
Bubble tea’s popularity thus becomes an avenue for soft power, helping shape international views of Taiwan
through an accessible and widely embraced cultural export.
Approximately 69\% of all posts in the dataset are in English,
suggesting that most of these posts are exposed to English-speaking audiences.

\subsection{Unsupervised Topic Modeling}
To investigate the key topics in the dataset, we utilized social network analysis, time series analysis,
and Latent Dirichlet Allocation (LDA) for a robustness check (Appendix~D).
We also counted the most common food types to identify prominent items, potential connections, and latent themes
within the hashtags and text sections of the posts that might affect the algorithm and engagement metrics
\citep{Blei2003LDA,Jelodar2019TopicModeling}.
This approach allowed for visualizations and robust insight checks into the semantic space,
where, at the word level, we identify key themes and their relationships.
Crucially, this process helps reveal clusters and patterns in which social or political ideas may intersect with Taiwanese food.

\subsection{Quantitative Statistical Analysis}
Beyond merely examining the data for common themes and their relation to Taiwan,
we performed statistical $t$-tests and Analysis of Variance (ANOVA) for mentions of Taiwan to determine whether the theme truly matters.
For these tests, we used multiple keywords for Taiwan---including ``Republic of China'' and ``Formosa,'' as listed in Appendix~C.

Additionally, we tested whether there is a differentiation between hashtags and text containing references to Taiwan
in relation to algorithmic and engagement metrics (views, likes, comments).
This approach provides a more calculated understanding of the data that is not immediately apparent organically.

Additionally, by using an attention funnel, we visualize the process by which posts on social media platforms like Meta
move from exposure (views) to likes and comments---the pathway of the posts investigated.

\subsection{Quantifying Culinary Diversity with Entropy}
To systematically compare the representation of food types across similar countries and evaluate their potential use in gastrodiplomacy,
we computed the Shannon entropy for food-type mentions in online discourse using relative frequencies extracted from Instagram datasets
over the same period as those analyzed for Taiwan.
Shannon entropy provides a quantitative measure of diversity frequently used in social science~\cite{chang2025liberals}. 
Higher values indicate a more even, heterogeneous spread among categories,
while lower values signify concentration around a few dominant items.

\section{Results}

\subsection{Bubble Tea Dominates Taiwanese Gastrodiplomacy}

\begin{table*}[t]
\centering
\small                                
\setlength{\tabcolsep}{6pt}           
\renewcommand{\arraystretch}{1.05}
\caption{Distribution of Food Hashtags on Instagram (2020--2024).}
\label{tab:food-distribution}
\begin{tabular}{@{}lrrrrr@{}}
\hline
\textbf{Food} & \textbf{Hashtags} & \textbf{Texts} & \textbf{Total} & \textbf{\%} & \textbf{Cum.\%} \\
\hline
Boba                    & 175{,}840 & 72{,}309 & 248{,}149 & 66.00\% & 66.00\% \\
Bubble Tea              & 59{,}213  & 19{,}047 & 78{,}260  & 20.81\% & 86.81\% \\
Milk Tea                & 21{,}674  & 12{,}368 & 34{,}042  & 9.05\%  & 95.86\% \\
Soup Dumplings          & 5{,}332   & 2{,}449  & 7{,}781   & 2.07\%  & 97.93\% \\
Bubble Milk Tea         & 2{,}260   & 383      & 2{,}643   & 0.70\%  & 98.64\% \\
Beef Noodle Soup        & 1{,}412   & 1{,}087  & 2{,}499   & 0.66\%  & 99.30\% \\
Popcorn Chicken         & 373       & 570      & 943       & 0.25\%  & 99.55\% \\
Braised Pork Rice       & 109       & 349      & 458       & 0.12\%  & 99.67\% \\
Taiwanese Fried Chicken & 118       & 175      & 293       & 0.08\%  & 99.75\% \\
Stinky Tofu             & 90        & 167      & 257       & 0.07\%  & 99.82\% \\
Gua Bao                 & 79        & 112      & 191       & 0.05\%  & 99.87\% \\
Scallion Pancakes       & 56        & 117      & 173       & 0.05\%  & 99.92\% \\
Pineapple Cake          & 65        & 56       & 121       & 0.03\%  & 99.95\% \\
Taiwanese Sausage       & 18        & 96       & 114       & 0.03\%  & 99.98\% \\
Oyster Omelette         & 25        & 52       & 77        & 0.02\%  & 100.00\% \\
\hline
\end{tabular}
\end{table*}

\begin{figure}[t!]
\centering
\includegraphics[width=\linewidth,
                 height=0.34\textheight, 
                 keepaspectratio]{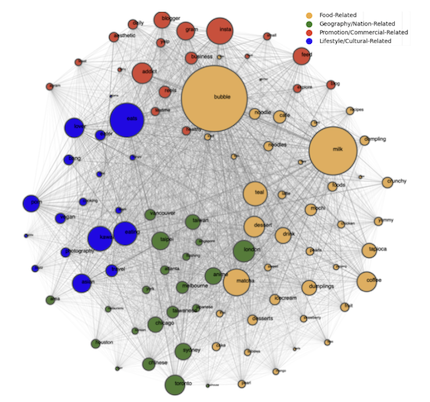}
\caption{Co-occurence of hashtags.}
\label{fig:cooccur}
\end{figure}

Table~\ref{tab:food-distribution} shows the distribution of food hashtags and their counts, sorted by frequency. Bubble tea is overwhelmingly dominant, appearing more than 95\% of the time. In contrast, soup dumplings occupy only 2\% and beef noodle soup takes up 0.66\%, which marks a precipitous drop from bubble tea and milk tea’s dominance. Notably, this distribution resembles a power-law distribution, which is common among social systems that exhibit the rich-get-richer phenomenon \citep{clauset2009powerlaw}.

Another way of visualizing the semantic space of online discourse around food is through a network of co-occurring words, shown in Figure~\ref{fig:cooccur}. The network map was generated using Term Frequency-Inverse Document Frequency (TF-IDF), filtering out stop words and words below the 90th percentile, and selecting the top 100 prominent words. Food-related words (orange) are the largest nodes.

Beyond food, the three other most significant categories are Lifestyle/Cultural-Related (blue) and Geography/Nation-Related (green). The most central nodes are hashtags related to Taiwan (i.e., ``Taiwan,'' ``Taipei,'' and ``Taiwanese''), which speaks to the centrality of Taiwan to bubble tea online. The appearance of other locations, such as ``Sydney'' and ``Toronto,'' and in the text network, ``Irvine'' and ``Westminster,'' speaks to the international presence of bubble tea in top international locations.

\subsection{Temporal Dynamics of Bubble Tea Attention}

\begin{figure*}[t]
\centering
\includegraphics[width=0.7\linewidth]{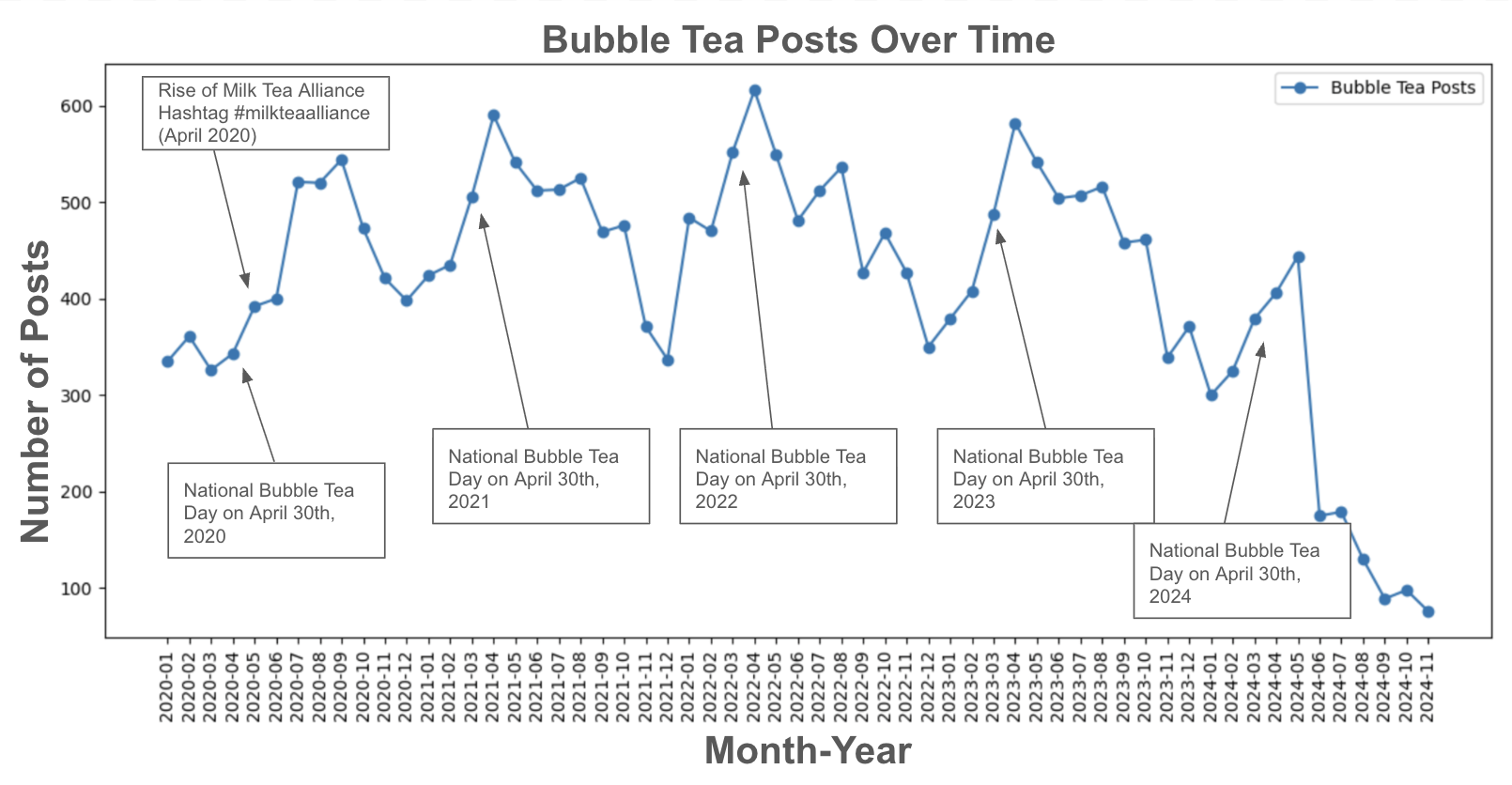}
\caption{Time series of bubble tea mentions on Instagram (2020--2024). 
Shaded regions indicate major socio-political events, including the 2020 Milk Tea Alliance 
and National Bubble Tea Day (April 30) peaks.}
\label{fig:timeseries_a}
\end{figure*}

Figure~\ref{fig:timeseries_a} shows the time series of bubble tea mentions over the past decade on Instagram. There is a clear seasonality in the attention given to bubble tea. While this may partly be due to increased consumption of cold drinks during the summer, the inciting factor is the Milk Tea Alliance in early 2020, which persisted throughout the summer as the Hong Kong protests continued. This suggests that broader socio-political events can drive heightened online attention to bubble tea. Additionally, peaks emerge annually around National Bubble Tea Day (April 30), an observance established by the Taiwanese-American chain Kung Fu Tea in 2018 \citep{kungfu2025}. These recurring spikes indicate a correlation between targeted promotions or social movements and an increase in bubble tea-related posts.

\begin{figure*}[t]
\centering
\includegraphics[width=0.7\linewidth]{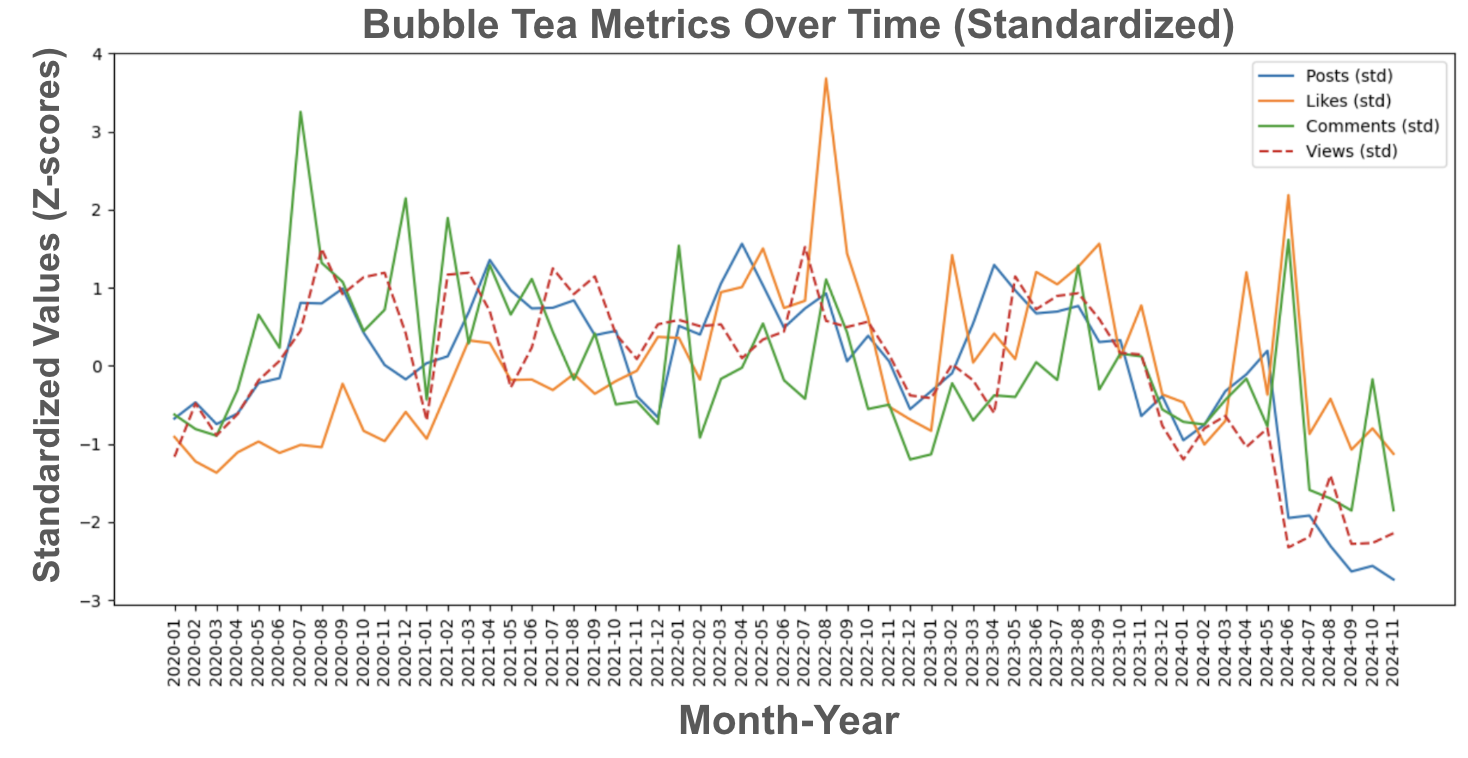}
\caption{Standardized time series of posts (blue), likes (orange), comments (green), and views (red) for bubble tea content on Instagram (2020--2024).}
\label{fig:timeseries_b}
\end{figure*}

Figure~\ref{fig:timeseries_b} shows the time series of bubble tea metrics, providing further evidence of the polarizing nature of the first summer with bubble tea. Specifically, it displays the standardized time series of posts (blue), likes (orange), comments (green), and views (red). Compared to all subsequent summers, the first summer features a significant bifurcation in comments and likes. In contrast, the summer of 2022 has many more likes relative to comments. The ratio between these two metrics matters: in social media studies, ``rationing'' refers to the phenomenon where the ratio of comments to likes can signal the level of engagement or controversy surrounding a post \citep{minot2021,riedl2023}. A higher comment-to-like ratio often indicates a more polarizing or contentious topic. Therefore, the elevated comments seen in 2020 likely reflect the broader socio-political tensions at the time—especially the Hong Kong protests and Tsai’s 2020 inauguration—prompting users to voice their opinions more actively than simply liking the posts.

\begin{figure*}[t]
\centering
\includegraphics[width=0.7\linewidth]{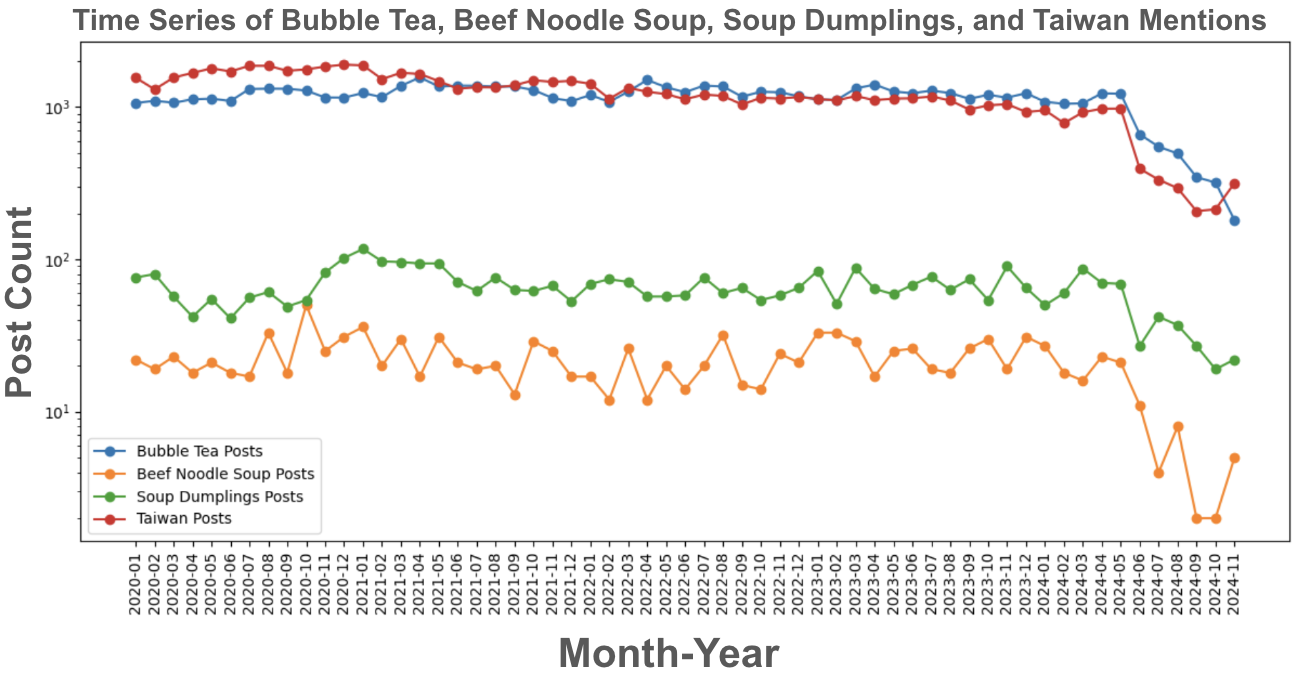}
\caption{Time series of Bubble Tea, Beef Noodle Soup, Soup Dumplings, and Taiwan mentions on Instagram (2020-2024).}
\label{fig:food_timeseries}
\end{figure*}

Figure~\ref{fig:food_timeseries} shows that bubble tea posts remain substantially more prevalent than both soup dumplings and beef noodle soup posts. This discrepancy highlights bubble tea's ongoing global appeal and cultural resonance, while soup dumplings---despite debates regarding their origins---have fewer mentions overall. Beef noodle soup stands out for its comparatively modest online presence, which suggests a smaller or more niche audience engaging with that topic. Notably, the trend lines reveal a strong correlation between bubble tea posts and Taiwan-related mentions. Bubble tea often serves as a cultural touchpoint, drawing attention to Taiwan and fueling broader discussions about its culinary heritage. However, the dominance of bubble tea content points to a concerning lack of diversity in the online discourse, as other dishes and cultural references remain relatively overshadowed. Furthermore, the apparent drop in content following Lai's inauguration suggests a shift in public focus or a decrease in online engagement with these topics. While it is unclear whether this decline is temporary or part of a more sustained trend, it raises questions about how political events might influence not only food-related conversations, but also how social media algorithms affect the visibility of a nation and its politics. This answers RQ1.

\subsection{Algorithmic and Engagement Measures are Diminished by Mentioning Taiwan}

Does inclusion of Taiwan increase or decrease viewership and engagement? When examining text mentions of Taiwan-related terms within bubble tea posts, the analysis shows a consistent pattern of lower engagement across all algorithmic and engagement metrics. Posts containing Taiwan-related text mentions received significantly fewer likes (\(t=-3.62,\, p=0.00030\)), views (\(t=-2.01,\, p=0.0445\)), and comments (\(t=-2.69,\, p=0.0070\)) compared to posts without such mentions. Specifically, the average number of likes for posts with Taiwan-related text mentions was 2{,}426.18, compared to 3{,}396.77 for posts without them. Posts with text mentions of Taiwan received an average of 134{,}910.96 views, compared to 170{,}359.83 views for posts without them. The average number of comments for posts with Taiwan-related text mentions was 42.20, while posts without such mentions received an average of 49.56 comments. We corroborate these findings with robustness tests using hashtags as well, found in Section~H in the SI, and demonstrate text and hashtags have no meaningful difference using ANOVA tests. Therefore, the data and subsequent analysis will not differentiate between whether ``Taiwan'' is mentioned via hashtag or within the text---it will be considered collectively based on the fact that the theme ``Taiwan'' is mentioned in either fashion.

In sum, mentions of Taiwan-related themes have a statistically significant negative effect on engagement; however, whether Taiwan is mentioned via text or hashtags is not substantial. Recalling the earlier proposition of bubble tea serving as a vehicle for Taiwanese soft power, we now face a paradox. Mentioning the cultural export of bubble tea along with its place of origin, Taiwan, is, put simply, doing more harm than good. This raises the question: Is this negative result due to organic audience disinterest, or is it a product of algorithmic biases?

\begin{figure}[t]
\centering
\includegraphics[width=\linewidth]{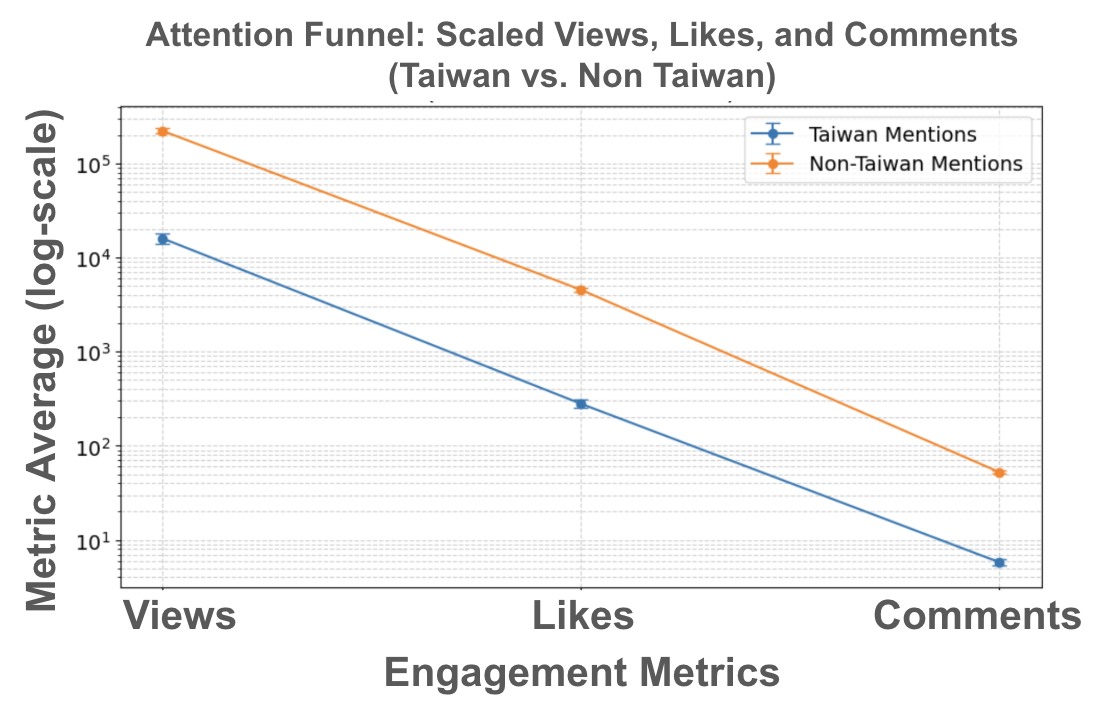}
\caption{Attention funnel comparing posts with and without Taiwan mentions. Posts referencing Taiwan show dramatically reduced exposure at the view stage.}
\label{fig:attention_funnel}
\end{figure}

To directly evaluate algorithmic bias, we consider aggregate levels of social media metrics across all levels of the attention funnel. Figure~\ref{fig:attention_funnel} shows whether mentioning Taiwan in bubble tea discourse is subject to algorithmic biases or organic audience disinterest, providing a comprehensive understanding of Taiwan’s bubble tea soft power digital visibility. The figure directly compares posts including Taiwan (blue) with posts without Taiwan (orange).

Figure~\ref{fig:attention_funnel} shows that non-Taiwan-themed posts have substantially higher average views than posts that mention Taiwan. This indicates that at the exposure level, the algorithm de-amplifies content related to Taiwan by a factor of approximately 12. However, once the attention funnel shifts to likes and comments, we observe a proportional drop in both metrics. There is not much significant difference, as both metrics slope downward at approximately the same scale. Although it is difficult to pinpoint the exact downstream effects of algorithmic de-amplification on relative engagement, we can say for certain that posts including Taiwan have significantly less exposure and engagement---fewer views, likes, and comments. This answers RQ2.

\subsection{Taiwan Has Low Levels of Gastrodiplomatic Diversity}

When considering the global context of this algorithmic suppression in views and its consequences, it is important to comparatively note Taiwan's regional neighbors and their gastrodiplomatic soft power, focusing on China, Thailand, and Japan.

China, Japan, and Thailand exhibit relatively high entropy scores (2.63, 2.37, and 2.33 bits, respectively), indicating that conversations about food in these contexts feature a wide array of dishes—no single food dominates the narrative. For example, China’s digital culinary landscape spans dumplings, hotpot, congee, and more, while Japan’s includes sushi, ramen, and curry. Such distributions suggest these cuisines have multifaceted digital footprints, reflective of both domestic culinary diversity and broad international appeal. By contrast, Taiwan’s entropy score is substantially lower at just 0.86 bits. The data show a steep concentration around a single item: ``boba'' or bubble tea. As visualized earlier, other iconic Taiwanese foods like beef noodle soup or guabao are nearly absent by comparison.

\begin{figure}[t]
\centering
\includegraphics[width=\linewidth]{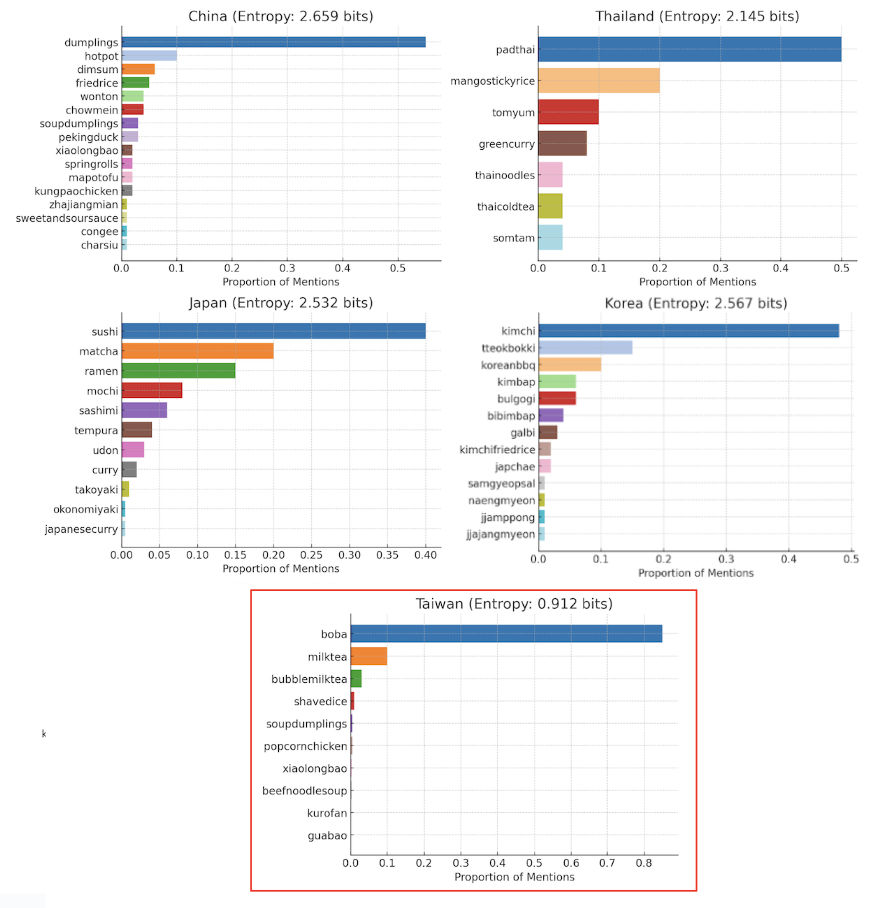}
\caption{Entropy Graphs for China, South Korea, Thailand, Japan, and Taiwan (Boxed in Red). Taiwan has least food diversity via entropy score.}
\label{fig:shap}
\end{figure}

This contrast in entropy scores has important implications for soft power and gastrodiplomacy. For China, South Korea, Japan, and Thailand, the diverse array of foods represented suggests that their international image is constructed around a tapestry of culinary exports. This diversity enables multiple cultural touchpoints and narratives, fostering greater global recognition of the national cuisine. In Taiwan’s case, the low entropy score demonstrates the global dominance of bubble tea as a cultural export. However, it also reveals a weakness: other elements of Taiwanese food culture are overshadowed or even rendered invisible in international discourse. While bubble tea’s ubiquity supports a potent, easily recognizable symbol for Taiwan, it may inadvertently restrict the richness and complexity of Taiwan’s broader cultural representation.

Together, these results illustrate how entropy can serve as a diagnostic tool to assess the strengths and limitations of a country's digital culinary identity. High entropy scores, as seen in China, Japan, and Thailand, reflect robust, diversified food narratives. Taiwan's sharply lower entropy highlights both the success and the risks of singular culinary branding in the algorithmic era.

\begin{table}[t]
\centering
\small
\setlength{\tabcolsep}{6pt}
\renewcommand{\arraystretch}{1.08}
\caption{Comparison countries and diminution in views after tagging country hashtags.}
\label{tab:comparison_views}
\begin{tabular}{@{}r l r l@{}}
\toprule
\textbf{Rank} & \textbf{Country} & \textbf{\% Change in Views} & \textbf{Food} \\
\midrule
1 & Thailand & $-38.7\%$ & Pad Thai \\
2 & Korea    & $-25.7\%$ & Kimchi \\
3 & China    & $-25.0\%$ & Dumplings \\
4 & Japan    & $-20.0\%$ & Sushi \\
5 & Taiwan   & $-9.0\%$  & Bubble Tea \\
\bottomrule
\end{tabular}
\end{table}

To assess how tagging a country affects content visibility for its most internationally recognized food, Table~\ref{tab:comparison_views} presents the percentage change in views after tagging posts with country-specific hashtags, such as ``\#taiwan'' or ``\#korea.'' For consistency, each country was paired with one iconic dish—Japan with sushi, Korea with kimchi, China with dumplings, Thailand with pad thai, and Taiwan with bubble tea—mirroring the method applied to Taiwan. This standardized approach allowed for a controlled comparison of how country tagging influences viewership across culinary contexts. While all five countries saw a decline in visibility, the severity varied significantly: Thailand experienced the steepest drop at $-38.7\%$, followed by Korea ($-25.7\%$), China ($-25.0\%$), Japan ($-20.0\%$), and Taiwan ($-9.0\%$). Although algorithmic suppression affected every case, the consequences for Taiwan are particularly concerning. Unlike its neighbors, Taiwan's global food identity is heavily concentrated around bubble tea, with few comparably prominent dishes. As a result, when this singular export suffers visibility penalties, Taiwan's entire digital food presence is disproportionately weakened.

\section{Discussion}

Empirically, our results demonstrate three key points. First, bubble tea dominates Taiwanese food content online, far outpacing all other dishes. Second, when bubble tea is explicitly associated with Taiwan, posts experience a dramatic decline in exposure---approximately a 91.7\% drop in views via the attention funnel---suggesting algorithmic suppression at the stage of initial visibility. Third, while likes and comments do not significantly differ once exposure is accounted for, the suppression of views at the outset cascades into reduced overall engagement. In short, algorithmic bias manifests most powerfully at the level of exposure, shaping downstream user interactions.

Taken together, our findings suggest that Taiwan's reliance on bubble tea as its primary vehicle for soft power leaves it especially vulnerable to platform-level biases. Whereas countries such as Japan, South Korea, and Thailand benefit from diverse culinary identities that can sustain their digital presence, Taiwan's cultural visibility is disproportionately concentrated in one product. When posts featuring bubble tea---the island's most globally recognized food---are penalized by Instagram's algorithms, Taiwan lacks alternative dishes to maintain its digital profile. Furthermore, entropy scores and visibility tests converge on a critical insight: Taiwan's digital gastrodiplomacy is structurally fragile, with its cultural reach narrowly constructed around a single symbol and particularly susceptible to algorithmic suppression.

This vulnerability has material as well as symbolic consequences. Although bubble tea originated in Taiwan and is framed as a national cultural export, recent reports indicate that Chinese and Southeast Asian companies increasingly dominate the industry due to economies of scale \citep{lo2023}. Thus, Taiwan risks the paradox of its signature soft power asset generating substantial commercial benefit for China---the very state it seeks to counterbalance diplomatically. Our analysis underscores how limited culinary diversification amplifies this exposure, particularly when compounded by algorithmic suppression of Taiwan-related content.

The broader implications extend beyond gastrodiplomacy. While social media has been lauded for its democratizing potential since the Arab Spring, subsequent events such as the 2016 election of Donald Trump have revealed its capacity to polarize and fragment societies. Against this backdrop, Instagram's demotion of content linking Taiwan with bubble tea raises pressing questions: What does it mean when one of the world's largest social media platforms suppresses the cultural expression of a democratic ally? If Taiwanese cuisine is consistently filtered out by automated moderation policies, Taiwan's international image risks being flattened, with bubble tea reduced to a depoliticized consumer good rather than a vehicle for cultural depth and sovereignty.

This paradox undermines the premise of bubble tea as an instrument of Taiwanese soft power. Rather than amplifying Taiwan's visibility, explicit mentions of the island diminish reach on Meta's platforms. One plausible explanation lies in Meta's 2024 policy shift to limit political content \citep{Treisman2024}. Taiwan---an inherently political topic---is likely captured by this filter, resulting in unintended consequences for its cultural diplomacy.

Algorithmic suppression not only weakens Taiwan's ability to leverage bubble tea as soft power but also raises ethical and political questions about platform governance. Future research should investigate which Taiwan-related terms most strongly trigger suppression, extend analysis to additional platforms such as X (formerly Twitter), and compare parallel cases (e.g., kimchi in South Korea or sushi in Japan) to assess whether algorithmic bias disproportionately affects certain national cuisines.

In sum, algorithmic design has direct consequences for cultural diplomacy. For Taiwan, whose most visible food export is tightly bound to questions of sovereignty, these dynamics mean that well-intentioned digital promotion may inadvertently undermine its soft power objectives. Addressing this tension---through diversification of cultural exports, advocacy for algorithmic transparency, or platform accountability---will be critical for sustaining Taiwan's visibility in the digital public sphere.

\bibliography{main}

\newpage

\appendix

\section{Section A: Keywords and Hashtags for Data Collection}
Following list derived from common Taiwanese food types and food hashtags for the search query in the data collection phase: 
{\raggedright\ttfamily
\#TaiwaneseFood, \#TaiwaneseCuisine, \#TaiwaneseEats, \#BubbleTea, \#BobaTea, \#BeefNoodleSoup, \#LuRouFan, \#TaiwanesePopcornChicken, \#TaiwaneseNightMarket, \#TaiwaneseDesserts, \#ShavedIce, \#MangoShavedIce, \#SoupDumplings.\par}

\section{Section B: Keywords for `Bubble Tea' Theme}
Following list is derived from common names for foods related to `Bubble Tea': \emph{Bubble Tea}, \emph{Boba}, \emph{Milk Tea}, \emph{Pearl Milk Tea}, \emph{Tapioca Tea}, \emph{Bubble Milk Tea}. This was used for finding posts related to the theme of bubble tea, be it in the text or hashtag form.

\section{Section C: Keywords for `Taiwan' Theme}
Following list is derived from common names for themes of `Taiwan': \emph{Taiwan}, \emph{Taiwanese}, \emph{Formosa}, \emph{Republic of China}, \emph{ROC}, \emph{Taipei}, \emph{New Taipei}, \emph{Taoyuan}, \emph{Hsinchu}, \emph{Miaoli}, \emph{Taichung}, \emph{Changhua}, \emph{Nantou}, \emph{Yunlin}, \emph{Chiayi}, \emph{Tainan}, \emph{Kaohsiung}, \emph{Pingtung}, \emph{Yilan}, \emph{Hualien}, \emph{Taitung}, \emph{Penghu}, \emph{Kinmen}, \emph{Matsu}. This was used for finding posts related to the theme of `Taiwan', be it in the text or hashtag form.

\section{Section D: Robustness Checks for Themes}
\subsection{Latent Dirichlet Allocation Testing (LDA)}
Aside from purely looking at the words present in the network maps for bubble tea-related posts and the themes they yield, topic distributions may be helpful to find hidden, latent connections.

\begin{table*}[t]
\centering
\begin{tabular}{@{}p{0.14\textwidth} p{0.82\textwidth}@{}}
\hline
\textbf{Text Topic Index} & \textbf{Text Topic Words} \\
\hline
0 & tea, milk, boba, sugar, brown, pearls, tapioca, matcha, drink, ice \\
1 & tea, boba, bubbletea, bubble, love, day, happy, time, new \\
2 & bubbletea, bubble, food, foodporn, foodie, tea, foodblogger, ipl, follow, instadaily \\
3 & boba, bubbletea, bobatea, eeeeats, tigersugar, bobalife, forkyah, kawaii, myfab, huffposttaste \\
4 & tea, bubble, boba, milk, drinks, try, new, bubbletea, sugar, food \\
5 & tag, giveaway, post, free, follow, pm, entries, like, th, winners \\
6 & boba, tea, delivered, order, menu, san, bobaful, delivery, home, kits \\
7 & boba, bobatea, bubbletea, milktea, bobamilktea, bubbletealover, bobalove, bobalife, bubbleteatime, bobatime \\
\hline
\end{tabular}
\caption{Text LDA topics ($K\!=\!8$): top words per topic.}
\label{tab:lda-text}
\end{table*}

\paragraph{LDA Themes for Text}
For texts (See Table 3), Topic 0 centers on the composition of bubble tea, emphasizing the drink’s ingredients and preparation, consistent with the network map’s food-related (orange) theme. Topics 1 and 7 shift toward bubble tea as an emotional and lifestyle-driven experience, with words such as ``love,'' ``happy,'' ``bubbletealover,'' and ``time'' framing it as more than just a beverage. This suggests a strong consumer-brand relationship, where bubble tea is associated with positive sentiment and personal moments, aligning with its frequent positioning as an indulgence or social activity, similar to the network map’s lifestyle/culturally-related (blue) theme. In Topics 3, 4, 5, and 6, bubble tea emerges as a key player in social media and influencer culture, corresponding to the Promotion/Commercial-Related topics in red.

\begin{table*}[t]
\centering
\begin{tabular}{@{}p{0.14\textwidth} p{0.82\textwidth}@{}}
\hline
\textbf{Hashtag Topic Index} & \textbf{Hashtag Topic Words} \\
\hline
0 & myfab5, bubbletea, sydneyfoodie, zomato, sydneyeats, asmr, sydneyfood, asmrfood, mukbang, sydney \\
1 & bubbletea, sharetea, pearl, bobatea, pearlmiktea, bubbletealover, pearlmiktea, pearltea, drinkup, bobaislife \\
2 & bubbletea, kawaii, illustration, boba, deliciousdrinks, cute, kawaiiaesthetic, digitalart, kawaiiart, art \\
3 & bubbletea, eeeeeats, boba, foodporn, forkyah, bobatea, foodgasm, buzzfeast, eater, bobalife \\
4 & bubbletea, boba, milktea, foodie, dessert, food, foodporn, instafood, tea, dailyfoodfeed \\
5 & bubbletea, boba, blogto, torontofood, bobatea, torontofoodie, instafood, torontoeats, markhamfood, bobacandle \\
6 & bubbletea, boba, foodblogger, bobatea, homecafe, yelpoc, yelpla, reels, happyhour, bubbleteasupply \\
7 & boba, bubbletea, bobatea, milktea, bubbletealover, bobalife, bobalove, bobatime, bobamilktea, bubbleteaaddict \\
\hline
\end{tabular}
\caption{Hashtag LDA topics ($K\!=\!8$): top words per topic (social/influencer vocabulary).}
\label{tab:lda-hashtags-b}
\end{table*}

\paragraph{LDA Themes for Hashtags}
For hashtags (See Table 4), Topic 0 highlights regional food trends and cultural positioning, with words like ``sydneyfoodie'' and ``sydneyeats'' reinforcing bubble tea’s embeddedness in specific geographic and digital food spaces, aligning with the Geography/Nation/Cultural-Related (green) theme in the network map. Topics 1 and 7 shift toward positioning bubble tea as an emotional and lifestyle-driven phenomenon, similar to the lifestyle/culturally-related (blue) theme in the network map. Topics 2, 3, 4, 5, and 6 showcase bubble tea’s roles in social media and influencer culture, paralleling the Promotion/Commercial-Related (red) theme from the network map.

Using LDA, we observed that its themes often do not mention Taiwan. Therefore, we investigated the importance of Taiwan mentions in either text or hashtags. While these mentions appear significant in the network maps, they are not necessarily dominant within the topic distributions generated by LDA, which focuses on explicit latent textual attribution. Instead, we performed statistical t-tests and ANOVA tests to ascertain and confirm the importance of the theme of Taiwan. The results proved that mentions of Taiwan significantly affect algorithmic and engagement metrics, as discussed in the paper.

\section{Section G: Statistical Tests for Mentions of Taiwan}
\subsection{T-Test For Hashtags}
The t-test results for hashtags also indicate a statistically significant difference across all key engagement metrics. Posts containing Taiwan-related hashtags received significantly fewer likes (\(t = -4.64,\ p = 3.59\times10^{-6}\)), views (\(t = -2.80,\ p = 0.0051\)), and comments (\(t = -7.93,\ p = 2.39\times10^{-15}\)) compared to posts without such hashtags. The average number of likes for posts with Taiwan-related hashtags was \(2{,}022.89\), compared to \(3{,}376.27\) for posts without them. Similarly, posts with Taiwan-related hashtags received an average of \(111{,}183.77\) views, compared to \(170{,}179.86\) views for posts without them. The average number of comments for posts with Taiwan-related hashtags was \(30.12\), while posts without such hashtags received an average of \(49.81\) comments.

These results suggest that the inclusion of Taiwan-related hashtags within bubble tea-themed posts has a particularly strong impact on comments, with a highly significant \(p\)-value of \(2.39\times10^{-15}\). This indicates that users are less likely to interact with posts mentioning Taiwan-related hashtags, which may reflect prior algorithmic influences or user perceptions—a dynamic that remains to be fully understood.

\subsection{Section H: No Difference between Mentioning of `Taiwan' via `Texts' versus `Hashtags': ANOVA Testing}
When examining engagement metrics in bubble tea-themed social media posts that include Taiwan-related terms, an ANOVA test was conducted to determine if there are significant differences in likes, views, and comments when Taiwan is mentioned in the `Hashtags' section versus the `Texts' section of a post. The analysis compared three groups: posts with both hashtags and text mentions of Taiwan, posts with only text mentions, and posts with no Taiwan-related mentions.

The ANOVA test results indicate that none of the differences in engagement metrics across the three groups are statistically significant. The test results for likes (\(F = 2.44,\ p = 0.0876\)), views (\(F = 1.24,\ p = 0.2896\)), and comments (\(F = 0.71,\ p = 0.4897\)) all show \(p\)-values above the 0.05 threshold for statistical significance. This suggests that the type of Taiwan-related mention—whether in the form of hashtags, text mentions, or both—does not have a meaningful impact on likes, views, or comments.

For likes, the greatest variance was observed between the groups, with an F-statistic of 2.44 and a p-value of 0.0876. Although this indicates some variation in likes between the groups, it is not enough to be considered statistically significant. Similarly, the p-values for views and comments show even less variance across the groups, suggesting that these metrics are relatively stable regardless of the type of mention.

These findings indicate that posts mentioning Taiwan-related terms in the text consistently experience lower engagement across likes, views, and comments. The differences are statistically significant in the t-tests, confirming that the observed lower engagement is unlikely to be due to chance. The ANOVA results confirm that the channel of mention (text vs. hashtag) does not materially alter this pattern.

\section{Section E: Visualizations for Power Law - See Figure 9}
\begin{figure}[t]
\centering
\includegraphics[width=\columnwidth]{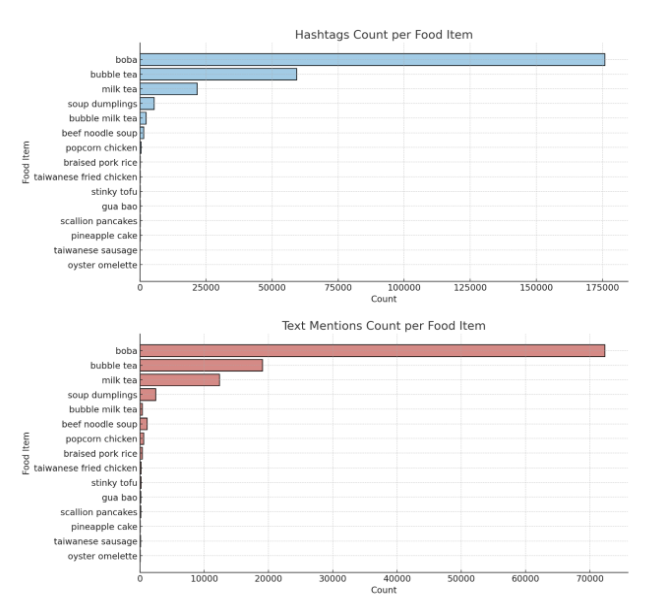}
\caption{Hashtag and text mention frequency distributions the theme of bubble tea in all its forms is mentioned the most via both texts and hashtags.}
\label{fig:powerlaw}
\end{figure}

\section{Section F: Semantic Network of Keywords - See Figure 10}
\begin{figure}[t]
\centering
\includegraphics[width=\columnwidth]{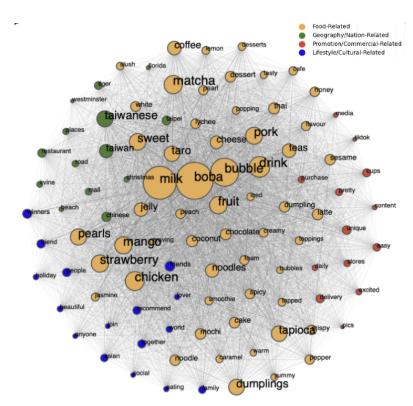}
\caption{Semantic network of keywords.}
\label{fig:keywords}
\end{figure}

\end{document}